\documentclass[prc,amsmath,amsfonts,amssymb,twocolumn,reprint]{revtex4-1}
\usepackage{caption}
\usepackage{subcaption}
\usepackage{graphicx}
\usepackage{bm}
\usepackage{color}
\usepackage[colorlinks=true,linkcolor=blue,citecolor=blue]{hyperref}

\begin{document}
\title{Combining symmetry collective states with coupled cluster theory: Lessons from the Agassi model Hamiltonian}
\date{\today}
\author{Matthew R.\ \surname{Hermes}}
\affiliation{Department of Chemistry, Rice University, Houston, TX, 77005, USA}
\author{Jorge \surname{Dukelsky}}
\affiliation{Instituto de Estructura de la Materia, CSIC, Serrano 123, E-28006 Madrid, Spain}
\author{Gustavo E.\ \surname{Scuseria}}
\affiliation{Department of Chemistry, Rice University, Houston, TX, 77005, USA}
\affiliation{Department of Physics and Astronomy, Rice University, Houston, TX, 77005, USA}

\begin{abstract}
The failures of single-reference coupled cluster for strongly correlated many-body systems is flagged at the mean-field
level by the spontaneous breaking of one or more physical symmetries of the Hamiltonian.
Restoring the symmetry of the mean-field determinant by projection reveals that coupled cluster
fails because it factorizes high-order excitation amplitudes incorrectly. However, symmetry-projected mean-field wave functions
do not account sufficiently for dynamic (or weak) correlation. Here we pursue a merger of symmetry projection and coupled
cluster theory, following previous work along these lines that utilized the simple Lipkin model system as a testbed
[J.\ Chem.\ Phys.\ \textbf{146}, 054110 (2017)].
We generalize the concept of a symmetry-projected mean-field wave function
to the concept of a symmetry projected state, in which the factorization of high-order excitation amplitudes in terms of low-order ones
is guided by symmetry projection and is not exponential, and combine them with coupled cluster theory in order to model the ground state
of the Agassi Hamiltonian. This model has two separate
channels of correlation and two separate physical symmetries which are broken under strong correlation.
We show how the combination of symmetry collective states and coupled cluster is effective in obtaining correlation energies and order parameters
of the Agassi model throughout its phase diagram.
\end{abstract}
\maketitle

\section{Introduction \label{sec_intro}}

In single-reference coupled cluster (CC) theory, high-order particle-hole excitation amplitudes out of
a reference determinant are factorized into products of lower-order particle-hole excitations \emph{via}
the exponential ansatz \cite{COESTER58, Cizek1966, KUMMEL78, ShavittBartlett}.
Single-reference CC methods are at the center of modern quantum chemistry calculations \cite{Bartlett2007}
because of their optimal combination of computational affordability and quantitative accuracy.
Similarly, there has been a renaissance of the method in nuclear physics where high precision
studies of medium-mass nuclei close to magic numbers were successfully performed in the last decade \cite{Dean, Gour2006, Hagen2014}.
However, single-reference CC is only accurate when applied to systems characterized by weak correlations,
a category that excludes systems exhibiting such important properties as superconductivity or superfluidity \cite{Henderson2014a, Duguet2016},
or nuclear deformation \cite{Duguet14},
and such ubiquitous
phenomena as bond breaking in quantum chemistry \cite{Bulik2015a}. Under weak correlation, the underlying reference wave function that coupled cluster usually takes as a given,
and which in single-reference CC must be a single determinant, is a qualitatively good approximation of the ground state eigenfunction.
Only if this condition is satisfied can the coupled cluster expansion of the wave function be truncated at a low order \cite{Bulik2015a}.

A single determinant labors under significant constraints in modeling real many-body wave functions.
In particular, when the correlations are strong, it is impossible for a single-determinantal
wave function to simultaneously exhibit accurate total energy and to respect the physical symmetries
of the Hamiltonian - that is, in order to get a qualitatively good total energy, a mean-field treatment
of the problem must break the real physical symmetries that the true many-body wave function would exhibit;\ this is
called the ``symmetry dilemma'' \cite{Lowdin1963}.  Coupled cluster methods can be built off of symmetry-broken determinants,
such as unrestricted, quasiparticle, or Bogoliubov coupled cluster methods \cite{Henderson2014a,Signoracci2015}.
These usually produce good total energies, but for finite systems, this comes
at the cost of other properties of the wave function, including
quantum numbers defined by physical symmetries such as angular momentum and particle number \cite{Sheikh2000,Cui2013}
which are lost in the broken-symmetry treatment.
In the thermodynamic limit, this is not an issue, but to obtain good properties for finite systems,
it is usually necessary to obtain explicitly symmetry-adapted
wave functions.

It is possible to restore good quantum numbers to a broken-symmetry mean-field wave function by projecting
out its symmetry-adapted component \cite{RingSchuck,BlaizotRipka, Schmid2004, Scus2011, Robledo2012,Jimenez-Hoyos2012}.
The resulting wave function is a linear combination of non-orthogonal determinants with good symmetries and
qualitatively good energies under strong correlation.
However, despite this qualitative superiority to single-reference CC, projected mean-field wave
functions are still not exact and a good deal of ``leftover'' dynamic (i.e.\ weak) correlation remains to be accounted for. Furthermore, the static/strong correlation included in the symmetry projection of a mean-field reduces to the broken-symmetry mean-field in the thermodynamic limit \cite{Jimenez-Hoyos2012}.
A formalism for restoring particle number in quasiparticle coupled cluster has been presented \cite{Duguet2016} but, to the best of our knowledge, not yet implemented.

We have recently explored the possibility of interpolating between projected mean-field and CCSD
for the attractive pairing (``reduced BCS'') model exhibiting mean-field number symmetry breaking
\cite{Degroote2016, Pittel},
as well as explicitly combining the two approximations for molecules exhibiting mean-field spin symmetry breaking \cite{Qiu2017}.
In a previous work \cite{Wahlen2017}, we used the Lipkin model \cite{Lipkin1965,Meshkov1965,Glick1965}
as a testbed to develop projected restricted coupled cluster (PRCC),
projected quasirestricted coupled cluster (PQCC), and projected unrestricted coupled cluster (PUCC). All three methods
combine mean-field symmetry breaking, symmetry-projection operators, and coupled cluster theory in various manners. They are described
in detail in Sec.\ \ref{sec_theory_formalisms} below.

Here we expand on this development using the Agassi model Hamiltonian \cite{Agassi1968,Davis1986} as a second testbed which has more elaborate physics.
The Agassi model is a schematic version of the paring-plus-quadrupole model \cite{Bes1969} extensively used in nuclear physics to
describe deformed superconducting nuclei. It combines the Lipkin model with the two-level pairing model \cite{HOGAASENFELDMAN1961}.
Although its Hamiltonian is not integrable, the number of states in its collective Hilbert space grows slowly with the
size of the system, as $O(n^2)$, and therefore the Hamiltonian can be diagonalized for systems of up to hundreds of particles.
However, despite its simplicity, the Agassi model is rich in non-trivial phenomena;\ it
has two separate physical symmetries that are broken at the mean-field level in various regions of its phase diagram \cite{Davis1986}.
We test two of our new methods (PRCC and PQCC)
on this model and find that they produce good agreement with exact results over the entire phase diagram.
The successful
tests of these methods on the Lipkin and now Agassi model Hamiltonian lay the foundation for their future application
to realistic finite many-body systems.
An Appendix details some algebraic properties and basis functions that are useful in performing calculations
on the Agassi model.

\section{Theory \label{sec_theory}}

\subsection{Symmetry-adapted formalisms \label{sec_theory_formalisms}}

The restricted coupled cluster (RCC) wave function is
\begin{eqnarray}
   |\mathrm{RCC}\rangle &=& e^{T} |0\rangle,
   \label{RCC_wvfn}
\end{eqnarray}
in which $|0\rangle$ is a symmetry-adapted determinant, usually obtained from a mean-field approximation
such as the restricted Hartree--Fock (RHF) method. The cluster operator, $T$,
consists of a sum of particle-hole excitations out the reference determinant, weighted by
excitation amplitudes (``$T$ amplitudes''), that do not change any
of the quantum numbers associated with  symmetries of the Hamiltonian. It can
be expanded, and the expansion can be truncated, in terms of the number of particle-hole excitations out of the reference determinant,
\begin{eqnarray}
   T &=& T_1 + T_2 + T_3 + \ldots.
   \label{T_expansion}
\end{eqnarray}
Because of the exponential ansatz [Eq.\ (\ref{RCC_wvfn})], high-order particle hole excitations arise both from
high-order terms in the expansion of the cluster operator and from products of lower-order terms.
For instance, triple excitations originate both from $T_3$ and from $T_1\times T_2$, in terms of
Eq.\ (\ref{T_expansion}).

In the standard RCC algorithm, the Hamiltonian is subject to a similarity transformation which renders it non-Hermitian,
$\bar{H}=e^{-T}He^{T}$, and the $T$ amplitudes are obtained by stipulating that the symmetry-adapted reference
is a right-hand eigenstate, $\bar{H}|0\rangle = E|0\rangle$. The similarity-transformed Schr\"{o}dinger equation is then left-projected
against excited states, leading to a set of nonlinear equations defining residuals, $\{R_1, R_2, R_3, \ldots\}$, which must be made to vanish by varying $T$
and which are coupled to each other \cite{Fan2006}.
That is, $R_n$ depends on at least $T_n$, $T_{n+1}$, and $T_{n+2}$, so that the equations for the residuals of each order must be solved simultaneously.
In practical applications this coupling chain is broken by setting amplitudes of an immediately higher order than a certain cutoff to zero;\ for
example, in RCC with double excitations (RCCD), $T_3$ and $T_4$ are assumed to vanish. In the regime of dynamic (weak) correlation,
this is a reasonable assumption and such methods give accurate energies. However, under the regime of static (strong) correlation,
$T_3$ and $T_4$ become non-negligible and RCCD breaks down;\ the nonlinear equations for $T$ frequently do not have solutions or lead to complex energies.
Alternatively, the cluster operator $T$ can be calculated using a variational method;\ this approach always yields real energies, but
under strong correlation variational RCC (vRCC) undercorrelates substantially \cite{Qiu2017}. Either way, RCC is not an appropriate
ansatz for systems characterized by strong correlation.

Strong correlation also corresponds to spontaneous symmetry breaking at the mean-field level. In other words,
a symmetry-adapted state such as $|0\rangle$ ceases to be the lowest-energy solution to the mean-field equations.
A lower-energy broken-symmetry determinant appears, which we label $|\Phi\rangle$ and which is related to the symmetry-adapted
state \emph{via} a Thouless transformation,
\begin{eqnarray}
   |\Phi\rangle &=& e^{T_1 + Q_1} |0\rangle,
   \label{Thouless}
\end{eqnarray}
omitting a normalization factor. 
Here, $Q_1$ consists of weighted (by ``$Q$ amplitudes'') single particle-hole excitations out of the symmetry-adapted reference which, unlike $T_1$,
\emph{change} the quantum numbers associated with unitary symmetries of the Hamiltonian.
Note that particle-hole excitations in the same basis, such as those in $T_1$ and $Q_1$, commute with one another.
An example of a symmetry-broken mean-field method is the spin-unrestricted Hartree--Fock (UHF) method
used in quantum chemistry. The molecular Hamiltonian used in quantum chemistry commutes with
the magnitude of the total electron spin angular momentum, $S^2$ and its $z$-axis component, $S_z$, and
the two corresponding quantum numbers are labeled $S$ and $M_S$, respectively. The UHF method allows for mean-field
symmetry breaking of the spin magnitude, $S$, but not $z$-axis component, $M_S$.
The corresponding $T_1$ and $Q_1$ operators are
\begin{eqnarray}
  T_1 &=& \sum_{i,a} t_i^a E_a^i,
  \label{T1_qchem}
  \\
  Q_1 &=& \sum_{i,a} q_i^a S_a^i,
  \label{Q1_qchem}
\end{eqnarray}
where $t_i^a$ and $q_i^a$ are the $T$ and $Q$ amplitudes and $E_a^i$ and $S_a^i$ are particle-hole excitations from the $i$th occupied
spatial orbital to the $a$th unoccupied spatial orbital. $E_a^i$ is an excitation which leaves both $S$ and $M_S$ unchanged,
while $S_a^i$ is an excitation that leaves $M_S$ unchanged but increases $S$ by 1 \cite{Qiu2016,Qiu2017}.
Combining these two excitations in the exponential ansatz [Eq.\ (\ref{Thouless})] results in a UHF state, $|\Phi\rangle$, which
(in general) contains multiple components with different $S$, so that $|\Phi\rangle$ itself cannot be described
by any single spin quantum number.

In the Agassi model we work with below, there are no single particle-hole excitations that preserve
the symmetries of the Hamiltonian, so $T_1$ is nonexistent. Therefore, for the rest of this work, we drop $T_1$.
The definition of the broken-symmetry determinant is restated as
\begin{eqnarray}
   |\Phi\rangle &=& e^{Q_1} |0\rangle.
   \label{UHF}
\end{eqnarray}

Since symmetry-broken determinants appear when RCC breaks down, a simple solution is to use the symmetry-broken
determinant itself as the reference for a coupled cluster expansion, resulting in the unrestricted coupled
cluster (UCC) wave function,
\begin{eqnarray}
   |\mathrm{UCC}\rangle &=& e^{U} |\Phi\rangle,
   \label{UCC_wvfn}
\end{eqnarray}
where $U$ consists of particle-hole excitations out of $|\Phi\rangle$ and, similar to $T$, is expanded and truncated
in terms of particle-hole excitation order,
\begin{eqnarray}
   U &=& U_1 + U_2 + U_3 + \ldots.
\end{eqnarray}
Neither $|\Phi\rangle$ nor the UCC wave function has good quantum numbers associated with symmetries,
and the individual terms in $U$ cannot easily be matched to specific transitions between symmetry
quantum numbers. Although total energies calculated with the UCC wave function are usually very accurate,
the results for wave function properties other than the energy are less reliable for finite systems
because of the loss of symmetry.

In a previous work \cite{Wahlen2017}, we experimented with a method that applies a symmetry projection
operator to a UCC wave function, which we named projected-unrestricted
CC (PUCC). The PUCC wave function is
\begin{eqnarray}
   |\mathrm{PUCC}\rangle &=& Pe^{U} |\Phi\rangle,
   \label{PUCC_wvfn}
\end{eqnarray}
where $P$ selects those components of the wave function with the desired
symmetry quantum numbers. PUCC was applied to the Lipkin model system \cite{Wahlen2017} and was found to improve on the accuracy of the total energies
calculated with UCC, especially in the limit of very strong correlation.
The Agassi Hamiltonian we will investigate below breaks number symmetry at the mean-field level under strong
correlation, and the corresponding ``unrestricted'' CC method is quasiparticle or Bogoliubov CC \cite{Henderson2014a,Signoracci2015}. 
Here, we do not further discuss UCC or PUCC, which will be presented in due time, and focus on the symmetry adapted variants discussed in detail below. 

An alternative to PUCC explored for the Lipkin model \cite{Wahlen2017}
is to substitute the symmetry-adapted RCC cluster operator, $T$, for $U$ in Eq.\ (\ref{PUCC_wvfn}).
The whole expression can then be written in the symmetry-adapted basis by using Eq.\ (\ref{UHF}) for the broken-symmetry
determinant, $|\Phi\rangle$. This is the projected-restricted CC (PRCC) wave function,
\begin{eqnarray}
   |\mathrm{PRCC}\rangle &=& Pe^{T+Q_1} |0\rangle.
   \label{PRCC_wvfn}
\end{eqnarray}
Note that the projection operator commutes with a symmetry-adapted excitation operator such as $T$, so that we can
move it to the right,
\begin{eqnarray}
   |\mathrm{PRCC}\rangle &=& e^T Pe^{Q_1}|0\rangle
   \nonumber \\ &=& e^T |\mathrm{PHF}\rangle,
   \label{PRCC_wvfn_2}
\end{eqnarray}
thus expressing PRCC as a coupled cluster ansatz with a symmetry-adapted cluster operator and a projected Hartree--Fock (PHF) wave
function as the reference ``determinant.''
However, a PHF state is actually a linear combination of several non-orthogonal determinants \cite{Jimenez-Hoyos2012}
which arise from applying the projection operator to the broken-symmetry mean-field determinant.
In the language of particle-hole excitations, PHF wave functions contain excitations
to all orders. To see how this comes about, we can closely examine the PHF wave function
\begin{eqnarray}
   |\mathrm{PHF}\rangle &=& Pe^{Q_1}|0\rangle
   \nonumber\\ &=& P\left(1+Q_1+\frac{Q_1^2}{2!} + \frac{Q_1^3}{3!} + \ldots\right)|0\rangle
   \nonumber\\ &=& F(K_2)|0\rangle,
   \label{PHF_wvfn}
\end{eqnarray}
where $K_2$ consists of symmetry-adapted double excitations which are products of single excitations in $Q_1$,
and $F$ is some polynomial of $K_2$ to all orders.
As discussed above, the powers of $Q_1$ arising
from the exponential creates components of the wave function that mixes the symmetry quantum numbers albeit in a ``controlled'' manner.
In selecting particular terms corresponding to the target quantum numbers from the exponential, the projection operator creates a new polynomial,
$F$, in which high-order particle-hole excitations are expressed in terms of products of lower-order ones by some
\emph{non-exponential} formula.
We have found in previous works that projecting out spin symmetry in this way leads to a hyperbolic sine
function \cite{Qiu2016},
and that projecting out number symmetry for the reduced BCS Hamiltonian results in a modified Bessel function of the first kind
 \cite{Degroote2016}. Attempts to model these functions with an exponential ansatz will fail
and this is at the root of why
single-reference restricted coupled cluster is unstable under strong correlation.

Another alternative to PUCC involves a straightforward generalization of Eq.\ (\ref{PRCC_wvfn}),
\begin{eqnarray}
   |\mathrm{PQCC}\rangle &=& Pe^{T+Q} |0\rangle,
   \label{PQCC_wvfn}
\end{eqnarray}
where $Q$ now contains $Q_1$, as in PRCC, but also can contain higher-order symmetry-quantum-number-switching particle-hole excitations,
\begin{eqnarray}
   Q &=& Q_1 + Q_2 + Q_3 + \ldots.
\end{eqnarray}
This was named the projected quasirestricted coupled cluster (PQCC) method in Ref.\ \onlinecite{Wahlen2017}.
PRCC can be characterized as a form of PQCC in which $Q$ is restricted to $Q_1$.
As in PRCC, each excitation in $Q$ can unambiguously be assigned a specific transition between symmetry quantum numbers,
because the particle-hole basis is symmetry-adapted. However, unlike PRCC, PQCC cannot generally be interpreted as coupled
cluster with a PHF reference state, because $Q$ is not limited to one-body excitations and $e^Q|0\rangle$ is therefore
not a Thouless transformation of the symmetry-adapted reference determinant. We use the name ``symmetry collective state,'' or ``SC,'' to
refer to this sort of generalization of PHF,
\begin{eqnarray}
   |\mathrm{SC}\rangle &=& P e^Q |0\rangle.
   \label{SC_general}
\end{eqnarray}
A PHF wave function is the simplest example of a symmetry collective state.
All symmetry collective states are symmetry-adapted
and contain particle-hole excitations to all orders out of a symmetry-adapted
reference determinant. PQCC can be described as the combination of a coupled cluster ansatz with a symmetry collective state
reference,
\begin{eqnarray}
   |\mathrm{PQCC}\rangle &=& e^T Pe^Q|0\rangle
   \nonumber \\ &=& e^T |\mathrm{SC}\rangle.
   \label{PQCC_wvfn_2}
\end{eqnarray}

In the test on the Lipkin model \cite{Wahlen2017}, 
PUCC was found to give superior accuracy in the strong-correlation limit, whereas PQCC was 
found to give better accuracy in the recoupling region, i.e., at correlation strengths near
where the lowest-energy mean-field wave function transitions between symmetry-adapted and symmetry-broken. 
Also, PQCC was found to have slightly superior accuracy in the energies compared
to PRCC at higher orders of truncation (i.e., triples or higher)
because of the additional excitations included in $Q$. 

Throughout the remainder of this work, we introduce a symbolic nomenclature in which terms all
of the wave functions hitherto described (other than UCC and PUCC) can be expressed:
\begin{eqnarray}
  |O_\mathrm{A}O_\mathrm{B}O_\mathrm{C}\ldots\rangle &\equiv& Pe^{O_\mathrm{A}+O_\mathrm{B}+O_\mathrm{C}+\ldots}|0\rangle,
  \label{nomenclature_def}
\end{eqnarray}
where $O_\mathrm{A}$, $O_\mathrm{B}$, and $O_\mathrm{C}$ are particle-hole excitation operators in the basis of the symmetry-adapted
reference state, $|0\rangle$. In this nomenclature, PHF, RCC, a general symmetry collective state, PRCC, and PQCC are
\begin{subequations}
\begin{eqnarray}
  |\text{PHF}\rangle &=& Pe^{Q_1}|0\rangle = |Q_1\rangle,
  \\
  |\text{RCC}\rangle &=& Pe^{T}|0\rangle = |T\rangle,
  \\
  |\text{SC}\rangle &=& Pe^{Q}|0\rangle = |Q\rangle,
  \\
  |\text{PRCC}\rangle &=& Pe^{T+Q_1}|0\rangle = |TQ_1\rangle,
  \\
  |\text{PQCC}\rangle &=& Pe^{T+Q}|0\rangle = |TQ\rangle.
\end{eqnarray}
\label{nomenclature_example}
\end{subequations}
Note that in the case of methods such as RCC, which include no
symmetry-breaking operators, the projection operator has no effect.

\subsection{Agassi Model \label{sec_theory_agassi}}

In this work, we apply the PRCC and PQCC methods to a more sophisticated testbed than the Lipkin model on which they were
originally developed:\ the Agassi model.
According to the labeling conventions of Davis and Heiss \cite{Davis1986},
the Agassi model describes a set of $2j$ spinless fermions occupying two sets of single-particle states, labeled $\sigma=-1$
and $\sigma=+1$, each of which is $2j$-fold degenerate. The individual states in each level are furthermore labeled with an index
$m$, with $-j \le m \le j$ and $m\neq 0$.
The Hamiltonian operator for this model is
\begin{eqnarray}
  H &=& \epsilon J_0 - \frac{V}{2} \left(J_+^2 + J_-^2\right) - g\sum_{\sigma,\sigma^\prime} A_\sigma^\dagger A_{\sigma^\prime},
  \label{Agassi_Hamiltonian}
\end{eqnarray}
where $\epsilon$, $g$, and $V$ are adjustable parameters and
where the one-body operators, $J$ and $A$, are
\begin{subequations}
\begin{eqnarray}
  J_0 &=& \frac{1}{2}\sum_m \left(c_{+1,m}^\dagger c_{+1,m}
       - c_{-1,m}^\dagger c_{-1,m}\right),
  \\
  J_+ &=& \sum_m c_{+1,m}^\dagger c_{-1,m},
  \\
  J_- &=& \sum_m c_{-1,m}^\dagger c_{+1,m},
  \\
  A_\sigma^\dagger &=& \sum_{m>0} c_{\sigma,m}^\dagger c_{\sigma,-m}^\dagger,
  \\
  A_\sigma         &=& \sum_{m>0} c_{\sigma,-m}        c_{\sigma,m}.
\end{eqnarray}
\label{Agassi_Operators_1}
\end{subequations}
That the size of the Hilbert space for this model grows quadratically with the number of particles, as mentioned in Sec.\ \ref{sec_intro}, is
demonstrated in the Appendix.

\begin{figure}[t]
\includegraphics[scale=0.5]{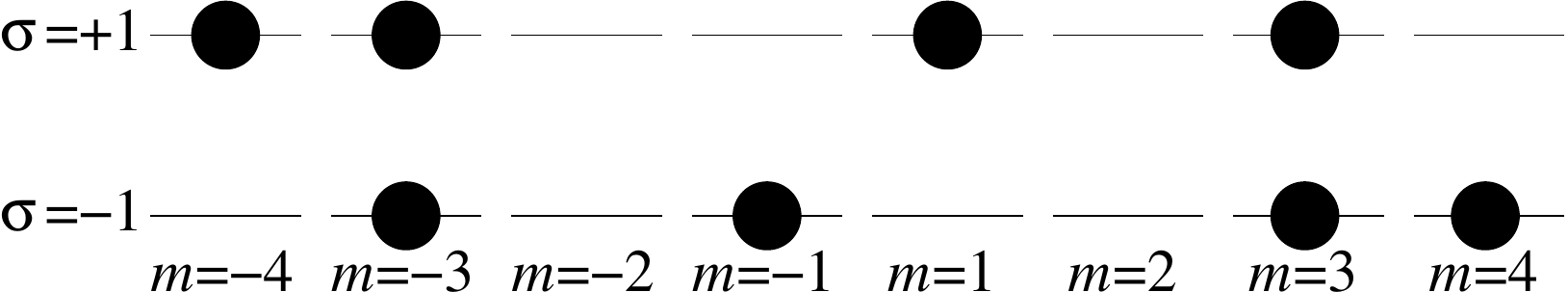}
\caption{A representative state of the Agassi model with $j=4$. Starting from the non-interacting ground state, in which all lower levels
and no upper levels are occupied [see Eq.\ (\ref{Zero_Ket})], this state is reached
by the action of two $J_+$ excitation operators at $m=-4$ and $m=1$, one
$A_{-1}$ annihilation operator at $m=2,-2$, and one $A_{+1}^\dagger$ creation operator at $m=3,-3$. \label{agassi_state_diagram}}\
\end{figure}

However, despite this simplicity, the physics of the Agassi model is rich in non-trivial phenomena.
The Agassi model is, in essence, a union of the Lipkin model \cite{Lipkin1965,Meshkov1965,Glick1965} and
the two-level pairing model \cite{HOGAASENFELDMAN1961}.
The Hamiltonian, Eq.\ (\ref{Agassi_Hamiltonian}), combines two qualitatively distinct channels of many-body correlation. The second term in
Eq.\ (\ref{Agassi_Hamiltonian}), involving $J_-$ and $J_+$ and with strength parameter $V$, is equal to the correlation term of the
Lipkin model Hamiltonian.
It moves pairs of particles in either the lower or upper level ``vertically'' up or down,
between $\sigma=-1$ and $\sigma=+1$, but cannot change the ``horizontal''
indices $m$. On the other hand, the third term in Eq.\ (\ref{Agassi_Hamiltonian}), involving $A_\sigma$ and $A_\sigma^\dagger$ and with strength
parameter $g$, is equivalent to the two-body term in the pairing Hamiltonian.
It moves pairs of particles with opposite $m$ indices (e.g.\ $\sigma,m$ and $\sigma,-m$)
to another pair of states at $\sigma^\prime,m^\prime$ and $\sigma^\prime,-m^\prime$.
A diagrammatic representation of all this is depicted in Fig.\ \ref{agassi_state_diagram}.

\begin{figure}[h!]
\includegraphics[scale=0.75]{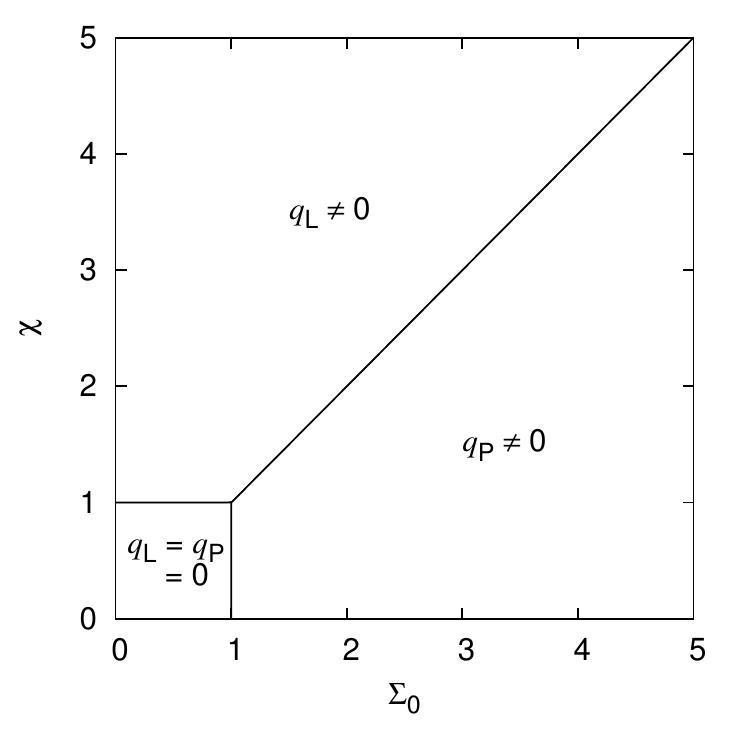}
\caption{The mean-field phase diagram of the Agassi model.
The two parameters $q_\mathrm{L}$ and $q_\mathrm{P}$ define the general
mean-field wave function of the Agassi model [Eqs.\ (\ref{UHF}) and (\ref{Q1})].
Along the ray where $\chi = \Sigma_0 > 1$, both $q_\mathrm{L}$ and $q_\mathrm{P}$
are nonzero.
\label{MF_phase_diagram}}\
\end{figure}

The two distinct channels of correlation correspond to the breaking of two separate symmetries of the Agassi Hamiltonian
at the mean-field level when the respective correlation strength parameters are large enough. The two relevant symmetries here
are parity, which is inherited from the Lipkin model and which breaks at the mean-field level when $V$
is large, and particle number, which breaks at the mean-field level when $g$ is large, as in the two-level pairing model \cite{HOGAASENFELDMAN1961}.
The parity symmetry operator, $\Pi$, and particle number operator, $N$, are
\begin{eqnarray}
  \Pi &=& e^{i\pi J_0},
  \label{parity_op}
  \\
  N &=& \sum_m \left( c_{+1,m}^\dagger c_{+1,m} + c_{-1,m}^\dagger c_{-1,m} \right),
  \label{number_op}
\end{eqnarray}
and the Hamiltonian has $[H,\Pi] = [H,N] = 0$. A state has $\Pi$-eigenvalue $+1$ or $-1$ depending
on whether it has an even or odd number of particle-hole excitations
from the lower level to the upper level and whether $j$ is even or odd.

The general mean-field wave function of the Agassi model which can break these two symmetries is given by
Eq.\ (\ref{UHF}) with 
\begin{eqnarray}
   Q_1 &=& q_\mathrm{L} J_+ + q_\mathrm{P} \left(A_{+1}^\dagger + A_{-1}\right),
   \label{Q1}
\end{eqnarray}
where the symmetry-adapted reference determinant is
\begin{eqnarray}
  |0\rangle &=& \left(A_{-1}^\dagger\right)^j |-\rangle,
  \label{Zero_Ket}
\end{eqnarray}
and where $|-\rangle$ is the physical vacuum. 
Eq.\ (\ref{Zero_Ket}) describes the full occupation
of the lower level of states ($\sigma=-1$) created by the maximum possible number of $A_{-1}^\dagger$ pair creation operators,
and the full vacancy of the upper level ($\sigma=+1$), indicated by the absence of any $A_{+1}^\dagger$ pair creation operators or
any $J_+$ particle-hole excitations.
In Eq.\ (\ref{Q1}), the two terms correspond to excitations along the two Agassi correlation channels and to breaking
of the two Agassi symmetries. Nonzero $q_\mathrm{L}$ corresponds to a parity-broken mean-field and nonzero $q_\mathrm{P}$ corresponds to a number-broken mean-field.
Throughout the rest of the paper, we refer to the symmetry-adapted reference state as the ``RHF'' state and the
general mean-field determinant $|\Phi\rangle$ as the ``UHF'' state which is equivalent of the Hartree-Fock-Bogoliubov vacuum of Ref.\ \onlinecite{Davis1986}.
Figure \ref{MF_phase_diagram} depicts the mean-field phase diagram;\
here and throughout the following we have made the change to dimensionless variables of Davis and Heiss \cite{Davis1986}
from $\epsilon$, $V$, and $g$ to
\begin{eqnarray}
   \chi &=& \frac{V(2j-1)}{\epsilon},
   \\
   \Sigma_0 &=& \frac{g(2j-1)+V}{\epsilon},
\end{eqnarray}
which are defined so that mean-field parity symmetry breaking occurs for $\chi > 1$ and mean-field number symmetry breaking occurs for $\Sigma_0 > 1$.

\begin{figure}[h!]
\includegraphics[scale=0.75]{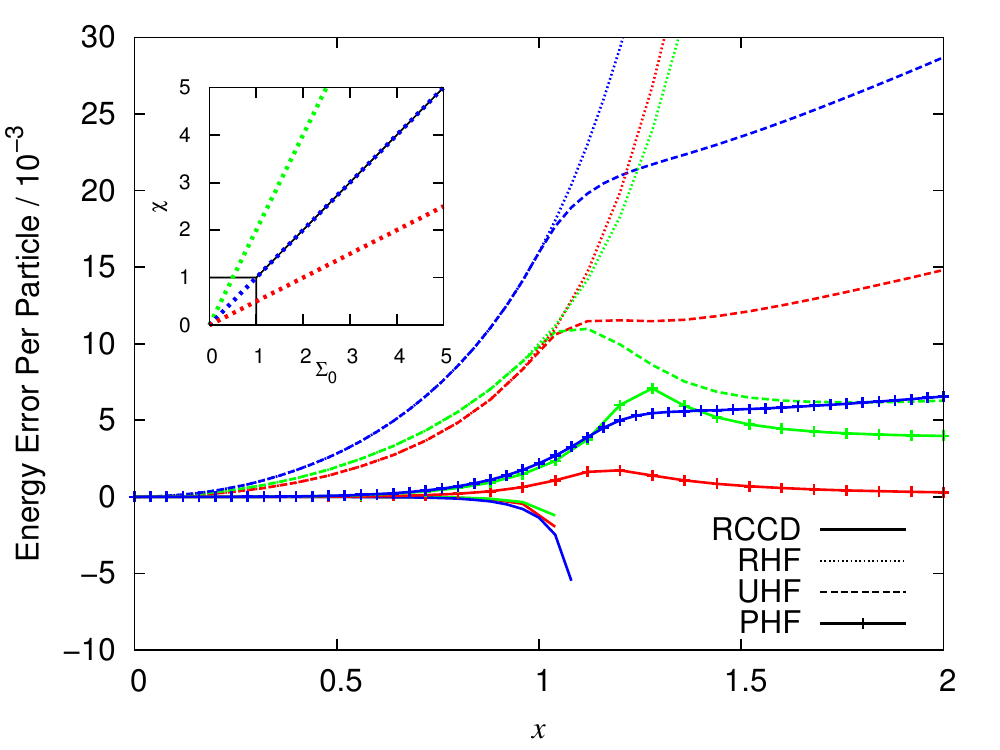}
\caption{
Ground-state energy errors
of the Agassi model with $j=20$
calculated by RCCD [Eqs.\ (\ref{RCC_wvfn}) and (\ref{T2})],
RHF,
[Eq.\ (\ref{Zero_Ket})]
UHF,
[Eqs.\ (\ref{UHF}) and (\ref{Q1})],
and PHF
[Eqs.\ (\ref{PHF_wvfn}) and (\ref{PHF_Agassi})].
The inset shows the lines through the phase diagram along which
the energies were calculated, and $x$ is the larger of $\chi$ or $\Sigma_0$.
No real solutions to the RCCD equations were found for $x>1.1$.
\label{RCCD_breakdown}}\
\end{figure}

The breaking of these symmetries at the mean-field level signals the impending failure of
typical single-reference methods to model the structure of the Agassi model
(or of any model Hamiltonian, for that matter).
Figure \ref{RCCD_breakdown} tracks this failure for the RCCD method
applied to an Agassi model with $j=20$. The RCCD wave function 
[$|T_2\rangle$ in the nomenclature of Eq.\ (\ref{nomenclature_def})]
is given by Eq.\ (\ref{RCC_wvfn})
with
\begin{eqnarray}
  T &=& T_2 = t_\mathrm{LL} J_+^2 + t_\mathrm{PP} A_{+1}^\dagger A_{-1}.
  \label{T2}
\end{eqnarray}
In Fig.\ \ref{RCCD_breakdown}, we obtain the amplitudes $t_\mathrm{LL}$ and $t_\mathrm{PP}$ through the standard CC algorithm
based on a similarity transformation;\ the existence of real solutions to the corresponding equations is not guaranteed.
Under weak correlation;\ i.e., $\chi < 1$ and $\Sigma_0 < 1$, the RCCD method is nearly exact. However,
near $\chi = 1$ or $\Sigma_0 = 1$, RCCD begins to overcorrelate, and starting shortly thereafter, we were unable
to find solutions to the nonlinear equations. Under strong correlation, $T_3$ and $T_4$ becomes non-negligible;\ RCCD calculations
assume that it is zero and thus fail.

The failure of restricted CC is due to its inability to model the strong correlation captured by the symmetry
breaking at $\chi = 1$ and $\Sigma_0 = 1$ in the mean-field wave function. Starting at $x = 1$ in Fig.\ \ref{RCCD_breakdown},
the UHF solution to the Agassi model begins to diverge from the RHF state. The energy errors of the
UHF solution are certainly much lower than the errors of the RHF energy. Going from UHF to PHF [$|Q_1\rangle$ in terms
of Eq.\ (\ref{nomenclature_def})]
further improves the energies by a significant margin. (A variation-after-projection method is used to obtain the PHF results
in Fig.\ \ref{RCCD_breakdown} and everywhere else in this work.) For the Agassi model the PHF wave function is explicitly
\begin{eqnarray}
  |\text{PHF}\rangle &=& |Q_1\rangle = Pe^{Q_1}|0\rangle 
  \nonumber\\ &=& P\sum_{l_1,l_2} \frac{q_\mathrm{L}^{l_1}}{l_1!}\frac{q_\mathrm{P}^{l_2}}{l_2!} J_+^{l_1}\left(A_{+1}^\dagger + A_{-1}\right)^{l_2}|0\rangle
  \nonumber\\ &=& \sum_{l_1,l_2} \frac{q_\mathrm{L}^{2l_1}}{(2l_1)!}\frac{q_\mathrm{P}^{2l_2}}{(l_2!)^2} J_+^{2l_1} \left(A_{+1}^\dagger A_{-1}\right)^{l_2} |0\rangle
  \nonumber\\ &=& \mathrm{cosh} (q_\mathrm{L} J_+) I_0 \left(2q_\mathrm{P}\sqrt{A_{+1}^\dagger A_{-1}}\right)|0\rangle,
  \label{PHF_Agassi}
\end{eqnarray}
where $Q_1$ is given by Eq.\ (\ref{Q1}) and
where $I_0$ is a modified Bessel function of the first kind.
The projection operator selects those terms with even powers
of $J_+$, which preserve parity symmetry,
and those terms with equal powers of $A_{+1}^\dagger$ and $A_{-1}$, which preserve
particle number. The resulting excitations are separable into a product of a
hyperbolic cosine of $J_+^2$
(as reported in the work on Lipkin parity projection in Ref.\ \onlinecite{Wahlen2017})
and a Bessel function of $A_{+1}^\dagger A_{-1}$
(as reported in the work on number projection in Ref.\ \onlinecite{Degroote2016}).

The PHF wave function is exact for the Agassi model in the strong attractive pairing limit ($\Sigma_0 \to \infty$ and $\Sigma_0 \gg \chi$)
and in the simultaneous thermodynamic and strong Lipkin correlation limit
($\chi \to \infty$, $\chi \gg \Sigma_0$, and $j \to \infty$). The behaviors of the red ($\Sigma_0 > \chi$) and
green ($\chi > \Sigma_0$) curves in Fig.\ \ref{RCCD_breakdown} are suggestive of these limits.
There is still visible error in the PHF energies, however, especially near the recoupling region.
We attribute this to weak (dynamical) correlation and therefore we expect that this error can be
substantially cured with our PRCC approach. Augmenting Eq.\ (\ref{PHF_Agassi})
with the doubles cluster operator of RCCD results in a PRCC wave function which, in the nomenclature of Eq.\ (\ref{nomenclature_def}), is
labeled $|T_2Q_1\rangle$:
\begin{eqnarray}
  |T_2Q_1\rangle &=& Pe^{T_2+Q_1}|0\rangle
  \nonumber\\ &=& e^{t_\mathrm{LL}J_+^2+t_\mathrm{PP}A_{+1}^\dagger A_{-1}} |\text{PHF}\rangle,
  \label{PRCCD_Agassi}
\end{eqnarray}
Note that, although the $|T_2Q_1\rangle$ wave function has more parameters than the PHF ($|Q_1\rangle$) wave function, the excitations
are still separable into a product of two polynomials, one for the $J_+^2$ excitations related to the Lipkin correlation channel and
one for the $A_{+1}^\dagger A_{-1}$ excitations related to the pairing correlation channel.

There is one limit of strong correlation in which PHF is not exact regardless of system size:\ the simultaneous
strong correlation limit along both channels, $\Sigma_0=\chi\to\infty$. The PHF energy error along that
line in Fig.\ \ref{RCCD_breakdown} (blue) grows monotonically with correlation strength instead
of converging, suggesting a non-dynamical correlation effect not captured by PHF in this region of the phase diagram.
We conjecture, and confirm in Sec.\ \ref{sec_results} below,
that the relatively poorer performance of PHF in this region is related to the lack of any terms
coupling correlation along the two channels. An analogy can be made to the difference between a
wave function obtained in an adiabatic approximation separating two different
interaction channels and one which includes non-adiabatic coupling. We anticipate that the latter wave function
will outperform the former in modeling a system in which the characteristic excitations
of the two channels have similar energy scales.

The lowest-order way to couple $J_+^2$ excitations to $A_{+1}^\dagger A_{-1}$ excitations is to 
include a higher-order $Q$ term:
\begin{eqnarray}
  Q_2 &=& q_\mathrm{LP} J_+ \left(A_{+1}^\dagger + A_{-1}\right).
  \label{PQCCSd_Q}
\end{eqnarray}
Although we call this $Q_2$, we note that a number-breaking, parity-preserving term with
$A_{+1}^\dagger A_{+1}^\dagger + A_{-1}A_{-1}$ as the operator part should formally be added to Eq.\ (\ref{PQCCSd_Q}).
We don't expect this latter excitation to be important, so we omit it here.
$Q_2$ as given by Eq.\ (\ref{PQCCSd_Q}) changes the quantum numbers of parity and number symmetry simultaneously.
Adding $Q_2$ to the excitations included in PHF generates a more general symmetry collective state labeled $|Q_1Q_2\rangle$:
\begin{widetext}
\begin{eqnarray}
  |Q_1Q_2\rangle &=& Pe^{Q_1+Q_2}|0\rangle
  \nonumber \\ &=& P\sum_{l_1,l_2,l_3} \frac{q_\mathrm{L}^{l_1}}{l_2!}\frac{q_\mathrm{P}^{l_2}}{l_2!}\frac{q_\mathrm{LP}^{l_3}}{l_3!}
  J_+^{l_1+l_3}\left(A_{+1}^\dagger + A_{-1}\right)^{l_2+l_3}|0\rangle
  \nonumber \\ &=& \sum_{l_1,l_2} \frac{(2l_2)!}{(l_2!)^2} \sum_{l_3=0}^{\mathrm{min}(2l_1,2l_2)}
  \left(
  \frac{q_\mathrm{L}^{2l_1-l_3}}{(2l_1-l_3)!}
  \frac{q_\mathrm{P}^{2l_2-l_3}}{(2l_2-l_3)!}
  \frac{q_\mathrm{LP}^{l_3}}{l_3!}
  \right)
  J_+^{2l_1} \left(A_{+1}^\dagger A_{-1}\right)^{l_2} |0\rangle.
  \label{PHFd}
\end{eqnarray}
\end{widetext}
Unlike PHF ($|Q_1\rangle$), $|Q_1Q_2\rangle$ is not a symmetry projection of a mean-field wave function,
because the exponential of $Q_2$ cannot be considered a Thouless transformation.
Making the same modification to the $|T_2Q_1\rangle$ wave function generates a PQCC method
labeled $|T_2Q_1Q_2\rangle$,
\begin{eqnarray}
  |T_2Q_1Q_2\rangle &=& Pe^{T_2+Q_1+Q_2}|0\rangle
  \nonumber \\ &=& e^{t_\mathrm{LL}J_+^2+t_\mathrm{PP}A_{+1}^\dagger A_{-1}}
  |Q_1Q_2\rangle.
  \label{PQCCSd}
\end{eqnarray}
We expect $|Q_1Q_2\rangle$ and $|T_2Q_1Q_2\rangle$ to outperform PHF ($|Q_1\rangle$)
and $|T_2Q_1\rangle$, respectively, in the region of the phase diagram where both correlation strength parameters
are simultaneously high.

Another potential way to couple the two correlation channels is in the cluster operator; i.e., by including terms such as
\begin{eqnarray}
  T_4 =& t_\mathrm{LLPP} J_+^2 A_{+1}^\dagger A_{-1} + \ldots,
\end{eqnarray}
which is similar to $Q_2$ in that it excites along both correlation channels simultaneously. However, in this work
we seek a method that can handle strong correlation at the doubles level or lower, and $T_4$ contains quadruple excitations.
Furthermore, CC methods excel at describing
weak, dynamic correlation, and the coupling between pairing and Lipkin channels is expected when correlation along both channels is strong.
For these reasons we do not here pursue methods using $T_4$.

\section{Results \label{sec_results}}

We test a PRCC wave function ($|T_2Q_1\rangle$) and a PQCC wave function ($|T_2Q_1Q_2\rangle$) on the Agassi model and examine both
the accuracy of the computed energies and the quality of the wave function, as measured by several order
parameters. We compare $|T_2Q_1\rangle$ and $|T_2Q_1Q_2\rangle$ to the results from exact diagonalization of the Agassi model
(``FCI'') as well as a variational implementation of RCCD ($|T_2\rangle$) and the PHF ($|Q_1\rangle$) and $|Q_1Q_2\rangle$ wave functions.
We take two separate approaches to determining
the $T$ and $Q$ amplitudes:\ first we test a variational formalism in which all amplitudes
are determined by minimizing the energy expectation value over a given wave function, and then we test
a similarity-transformed basis approach more common to coupled cluster theory. The former is robust
and has guaranteed solutions for any point on the phase diagram (although obtaining those solutions
may not always be computationally easy), whereas the latter is more practical for large systems
where FCI codes are not available. We continue to use the $j=20$ system which is large enough to significantly
reduce finite-size effects.

\subsection{Variational Energies \label{sec_results_Ev}}

\begin{figure}
\begin{subfigure}[b]{0.5\textwidth}
\includegraphics[scale=1]{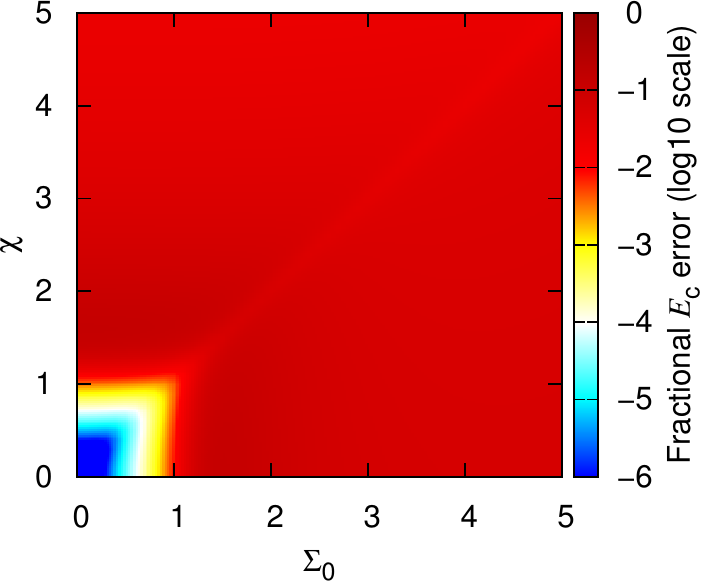}
\caption{RCCD ($|T_2\rangle$) wave function.
\label{EvRCCD}}\
\end{subfigure}

\begin{subfigure}[b]{0.5\textwidth}
\includegraphics[scale=1]{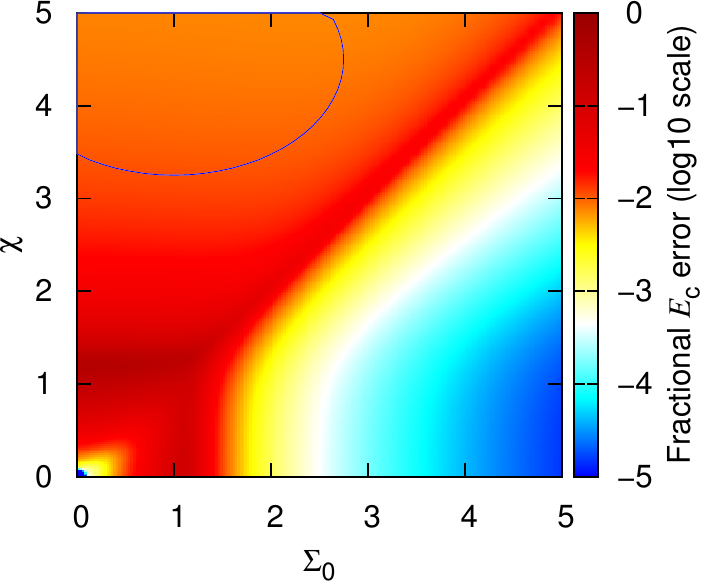}
\caption{PHF ($|Q_1\rangle$) wave function. The color axis is slightly shifted and an ellipse inscribed
to guide the eye towards the region of the phase diagram where $|Q_1\rangle$ results differ
from $|Q_1Q_2\rangle$ results.
\label{EPHF}}\
\end{subfigure}

\begin{subfigure}[b]{0.5\textwidth}
\includegraphics[scale=1]{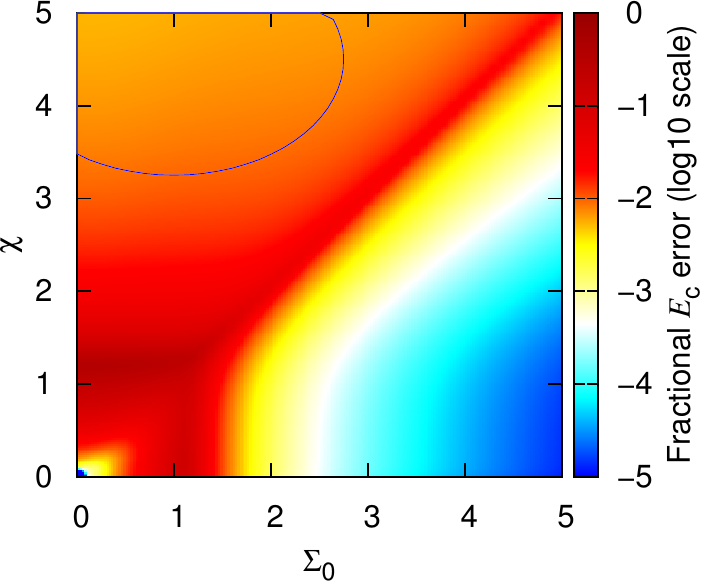}
\caption{$|Q_1Q_2\rangle$ wave function. The color axis is slightly shifted and an ellipse inscribed
to guide the eye towards the region of the phase diagram where $|Q_1\rangle$ results differ
from $|Q_1Q_2\rangle$ results.
\label{EPHFd}}\
\end{subfigure}
\caption{
The fractional correlation energy ($E_c$) error from FCI for an Agassi model with 40 particles, calculated
variationally
with the RCCD ($|T_2\rangle$), PHF ($|Q_1\rangle$), and $|Q_1Q_2\rangle$ wave functions.
\label{Ev1}}\
\end{figure}

Our variational treatments of CC theories are here denoted by the prefix ``v''. We make a Hermitian energy
expectation value stationary with respect to all wave function parameters simultaneously. For instance, the vRCCD
energy is obtained by solving the set of equations,
\begin{eqnarray}
  E_\text{vRCCD} &=& \frac{\langle T_2|H|T_2\rangle}{\langle T_2|T_2\rangle},
  \\
  0 &=& \frac{dE}{dt_\mathrm{LL}} = \frac{dE}{dt_\mathrm{PP}}.
\end{eqnarray}
In this section we evaluate the accuracy of various wave functions in recovering the correlation energy.
``Correlation energy'' in this case is defined with relation to the RHF energy:
\begin{eqnarray}
  E_c &=& E-\langle 0|H|0\rangle.
\end{eqnarray}
RHF is used as a reference point instead of the symmetry-broken mean field because the latter happens to give exact energies
in various limits of the Agassi model \cite{Henderson2014a,Wahlen2017},
limiting its usefulness as a reference in measuring the accuracy of approximate methods.

The error from FCI of the vRCCD ($|T_2\rangle$ wave function) 
correlation energy across the phase diagram is plotted in Fig.\ \ref{EvRCCD}.
The vRCCD method fails
quantitatively outside of the weak-correlation region at low $\chi$ and low $\Sigma_0$;\ at higher correlation
strengths vRCCD systematically undercorrelates by 5-10\%. There is no failure to converge the equations, as depicted in
Fig.\ \ref{RCCD_breakdown}, but that is because of the inherent robustness of the variational approach, as contrasted
to the projective method more commonly used in coupled cluster calculations \cite{ShavittBartlett}
and which RCCD (without the ``v'') utilizes. The vRCCD results still confirm that coupled cluster is poorly suited
to modeling strong correlation.

In contrast, the correlation energy error for PHF ($|Q_1\rangle$ wave function)
is plotted in Fig.\ \ref{EPHF}. The PHF energy has notably poorer
accuracy than the vRCCD energy in the weakly-correlated region of the phase diagram. However, at either strong correlation
limit, PHF outperforms vRCCD, undercorrelating by only up to 1\% in the strong Lipkin correlation region ($\chi > \Sigma_0$)
and achieving accuracy to within 0.01\% of the exact correlation energy in the strong pairing region ($\Sigma_0 > \chi$).
The better performance
in the strong pairing limit is attributable to the fact that PHF reduces to the PBCS wave function in that limit,
and in the strong-correlation limit of the attractive pairing model the PBCS wave function is exact \cite{Richardson1977,Roman2002,Degroote2016}.
The limit of the PHF wave function under strong Lipkin correlation, meanwhile, is also exact for the Lipkin model,
but only in the simultaneous thermodynamic and strong-correlation limits. The worse performance of PHF in the strong
Lipkin region compared to the strong pairing region is thus attributable to finite-size error.

Surprisingly, the $|Q_1Q_2\rangle$ wave function does not appear to improve significantly on the PHF results;\ indeed,
the difference between the two in Figs.\ \ref{EPHF} and \ref{EPHFd} is barely visible. Because these are variational
methods and because $|Q_1Q_2\rangle$ has one more parameter than does $|Q_1\rangle$, the energy of the $|Q_1Q_2\rangle$
wave function is indeed lower
than that of PHF, but not by a significant amount. This seems to falsify
our hypothesis discussed in Sec.\ \ref{sec_theory_agassi} that $|Q_1Q_2\rangle$ would make a critical improvement in the region
where both correlation strengths are simultaneously large. The only region of the phase diagram where any improvement
over PHF is visible is the strong-Lipkin region that, as discussed above, is subject to finite-size error.

\begin{figure}
\begin{subfigure}[b]{0.5\textwidth}
\includegraphics[scale=1]{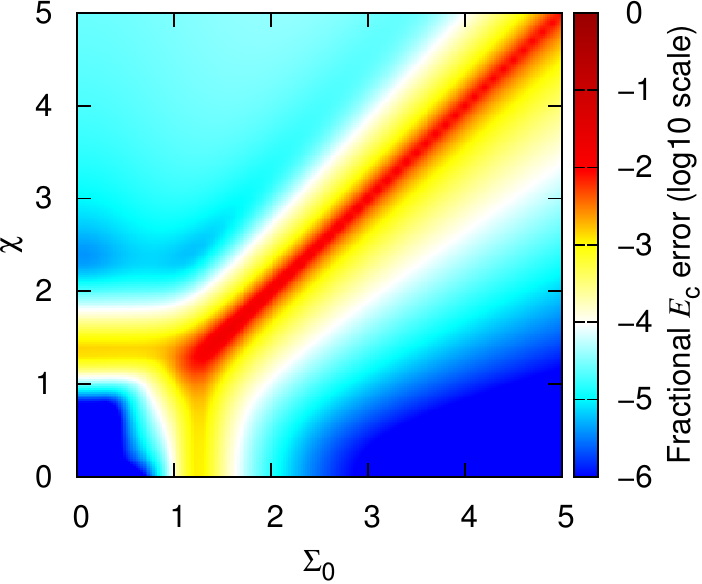}
\caption{PRCC ($|T_2Q_1\rangle$) wave function.
\label{EvPRCCD}}\
\end{subfigure}

\begin{subfigure}[b]{0.5\textwidth}
\includegraphics[scale=1]{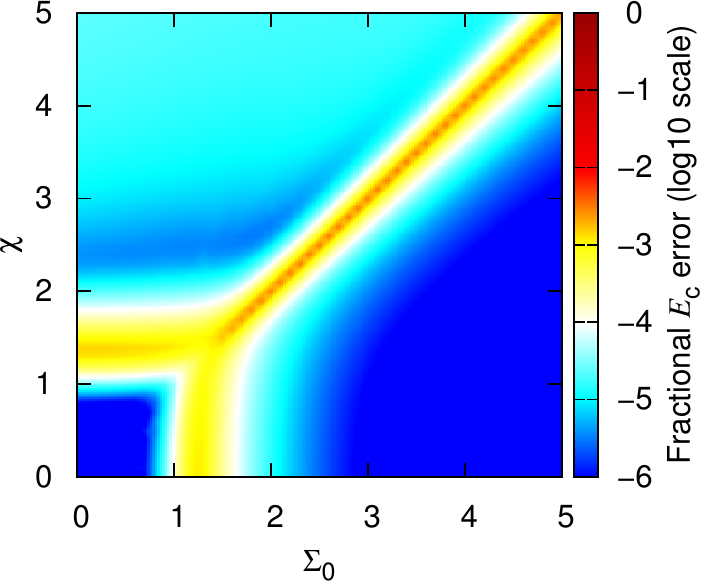}
\caption{PQCC ($|T_2Q_1Q_2\rangle$) wave function.
\label{EvPQCCSd}}\
\end{subfigure}
\caption{
The fractional correlation energy ($E_c$) error from FCI for an Agassi model with 40 particles, calculated
variationally
with the PRCC ($|T_2Q_1\rangle$) and PQCC ($|T_2Q_1Q_2\rangle$) wave functions.
\label{Ev2}}\
\end{figure}

Regardless, none of these three methods alone provide good-quality correlation energies across the entire phase diagram.
However, they complement one another when combined. The energy error of vPRCC ($|T_2Q_1\rangle$), which is a combination of
PHF and vRCCD, is plotted in Fig.\ \ref{EvPRCCD}.
The vPRCC energy is uniformly superior to both vRCCD and PHF, which is perhaps unsurprising since these are
all variational methods and the $|T_2Q_1\rangle$ wave function has the greatest number of parameters of these three.
However, the details are
significant:\ the vPRCC energy is nearly exact everywhere except in the regions near phase transitions.
The worst result is along the diagonal ray $\chi=\Sigma_0>1$, where vPRCC undercorrelates 1-2\%.
In this region of the phase diagram, both number and parity symmetry are broken at the mean-field level,
neither correlation channel dominates the other, and an adiabatic-like ansatz in which the excitations are
factorized into products of excitations along each channel is inappropriate.

Going from vPRCC to vPQCC (that is, going from $|T_2Q_1\rangle$ to $|T_2Q_1Q_2\rangle$) improves the quality
of the correlation energy in this region by an order of magnitude, as also shown in Fig.\ \ref{EvPRCCD}.
Now the hypothesis we discussed in Sec.\ \ref{sec_theory_agassi}, that $Q_2$
is critical to include when both correlation strengths are large, is confirmed. Clearly, the basic reason for this improvement is that
$Q_2$, which is included in the PQCC ($|T_2Q_1Q_2\rangle$) and $|Q_1Q_2\rangle$ wave functions but not the PRCC ($|T_2Q_1\rangle$) or RCCD ($|T_2\rangle$)
wave functions, couples Lipkin-like
excitations to pairing-like excitations at the level of the symmetry collective state, which is the part
of our wave function that is designed to account for strong-correlation effects. However, it does seem that this effect
only appears when combined with other double excitations, as in $|T_2Q_1Q_2\rangle$, and not when it is the sole double term,
as in $|Q_1Q_2\rangle$.

\subsection{Quality of the variational wave function \label{sec_results_wvfn}}

While PRCC ($|T_2Q_1\rangle$) and PQCC ($|T_2Q_1Q_2\rangle$) yield highly accurate energies, it remains to see whether those wave functions are accurate
for properties other than energies. We seek to test the accuracy of the Agassi wave function by probing several
order parameters derived from the wave function. The simplest is a measure of the average level of excitation;\ that is,
a measure of the number of particles in the upper level ($\sigma = +1$) on average:
\begin{eqnarray}
  n &=& \frac{1}{2j-1}\sum_m\langle c_{+1,m}^\dagger c_{+1,m}\rangle
  \nonumber \\ &=& \frac{1}{2j-1}\sum_m\langle N_{+1}\rangle.
  \label{Norder}
\end{eqnarray}
This is a straightforward measure of the strength of correlation in the wave function and does not distinguish between
Lipkin-channel and pairing-channel correlation:\ the closer our wave function is to RHF, the closer $n$ will be
to zero.

We additionally probe order parameters which track the effects of the two correlation channels independently.
For mean-field wave functions that are allowed to break symmetry, the obvious candidates involve expectation values
of the excitation operators that produce the broken-symmetry mean field:
\begin{eqnarray}
  J &=& \frac{\langle J_+\rangle}{2j-1},
  \label{Jorder}
  \\
  \Delta &=& \frac{\epsilon\Sigma_0-V}{2j-1}\left(\left\langle A_{+1}^\dagger\right\rangle + \left\langle A_{-1}^\dagger\right\rangle\right).
  \label{Dorder}
\end{eqnarray}
Note that $\Delta$ is the gap in the BCS sense \cite{RingSchuck}. However, these parameters cannot be used as written because
the approximate wave functions we are using (not to mention the exact wave function) are symmetry-adapted, so $J$ and $\Delta$
as defined in Eqs.\ (\ref{Jorder}) and (\ref{Dorder}) are zero by construction.
Therefore, for the purpose of tracking correlation strength in these wave functions, we modify Eqs.\ (\ref{Jorder})
and (\ref{Dorder}) to read
\begin{eqnarray}
  J &=& \frac{\sqrt{(\langle J_+^2\rangle + \langle J_-^2\rangle)/2}}{2j-1},
  \label{Jorder2}
  \\
  \Delta &=& \frac{\epsilon\Sigma_0-V}{2j-1}\left(
  \sqrt{\left\langle A_{+1}^\dagger A_{+1}\right\rangle} + \sqrt{\left\langle A_{-1}^\dagger A_{-1}\right\rangle}
  \right),
  \label{Dorder2}
\end{eqnarray}
which, if evaluated with a UHF wave function, give the same results as Eqs.\ (\ref{Jorder}) and (\ref{Dorder}) in the thermodynamic limit.

\begin{figure}
\begin{subfigure}[b]{0.5\textwidth}
\includegraphics[scale=1]{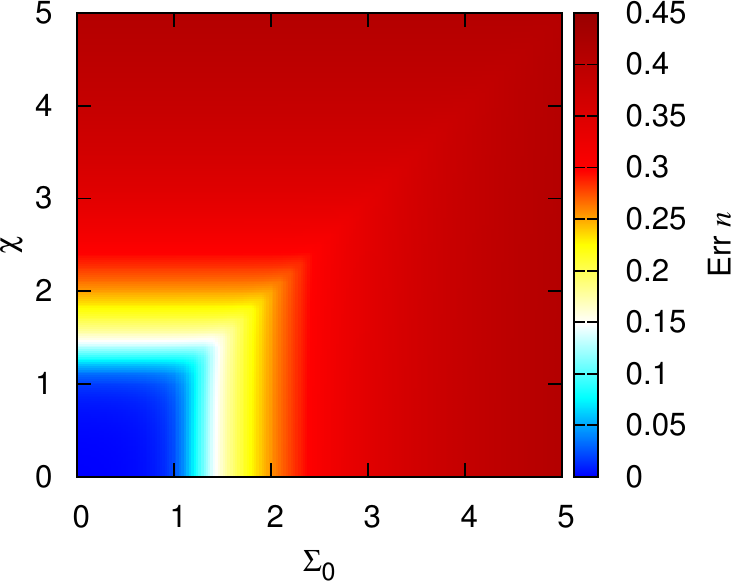}
\caption{Order parameter $n$ [Eq.\ (\ref{Norder})]
\label{OrderParam_exact_n}}\
\end{subfigure}

\begin{subfigure}[b]{0.5\textwidth}
\includegraphics[scale=1]{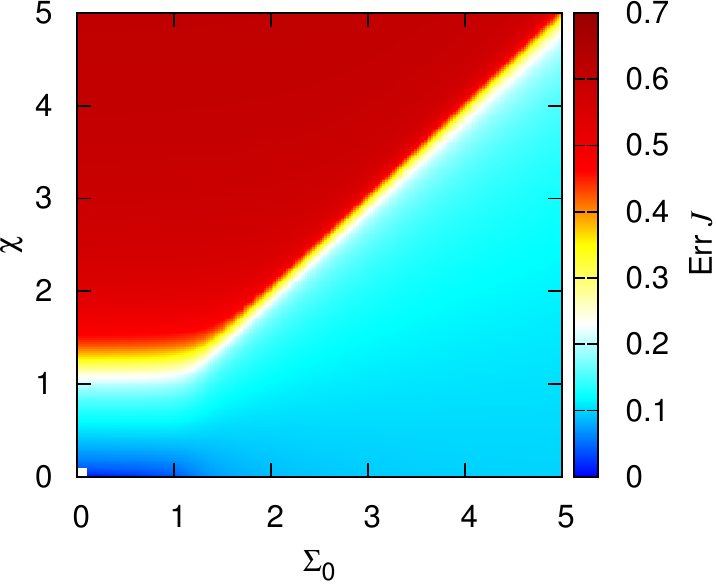}
\caption{Order parameter $J$ [Eq.\ (\ref{Jorder2})]
\label{OrderParam_exact_J}}\
\end{subfigure}

\begin{subfigure}[b]{0.5\textwidth}
\includegraphics[scale=1]{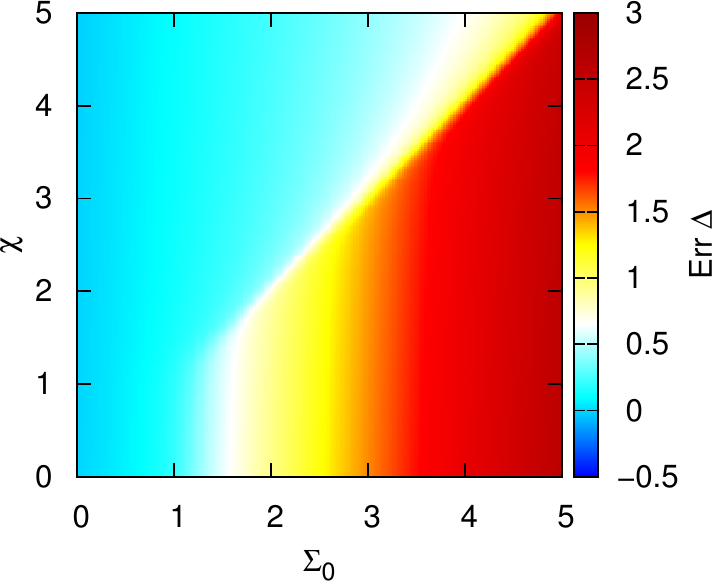}
\caption{Order parameter $\Delta$ [Eq.\ (\ref{Dorder2})]
\label{OrderParam_exact_D}}\
\end{subfigure}
\caption{Three order parameters
for the 40-particle Agassi model calculated from the FCI wave function.
\label{OrderParam_exact}}\
\end{figure}

Figure \ref{OrderParam_exact} plots the values of $n$, $J$, and $\Delta$ calculated using Eqs.\ (\ref{Norder}),
(\ref{Jorder2}), and (\ref{Dorder2}) from the exact (FCI) wave function. It can be seen that $n$ is large whenever
the system is strongly correlated along either channel, whereas $J$ and $\Delta$ are large only when
Lipkin correlation or pairing correlation, respectively, is strong.

\begin{figure}
\begin{subfigure}[b]{0.5\textwidth}
\includegraphics[scale=0.75]{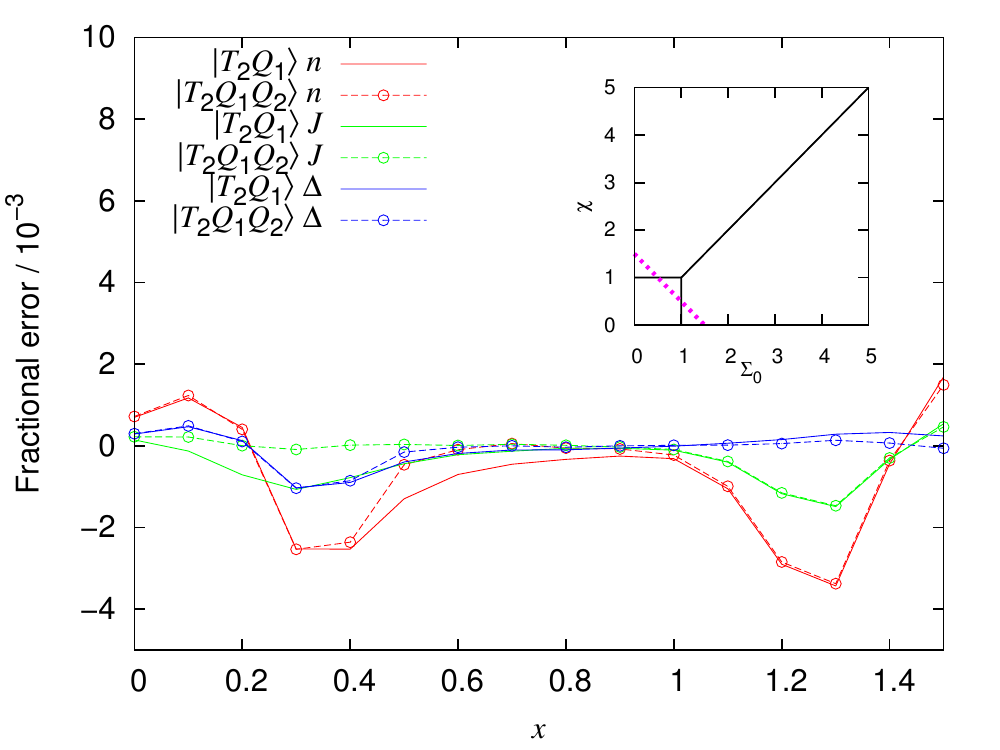}
\caption{$a=1.5$
\label{OrderParam_error15}}\
\end{subfigure}

\begin{subfigure}[b]{0.5\textwidth}
\includegraphics[scale=0.75]{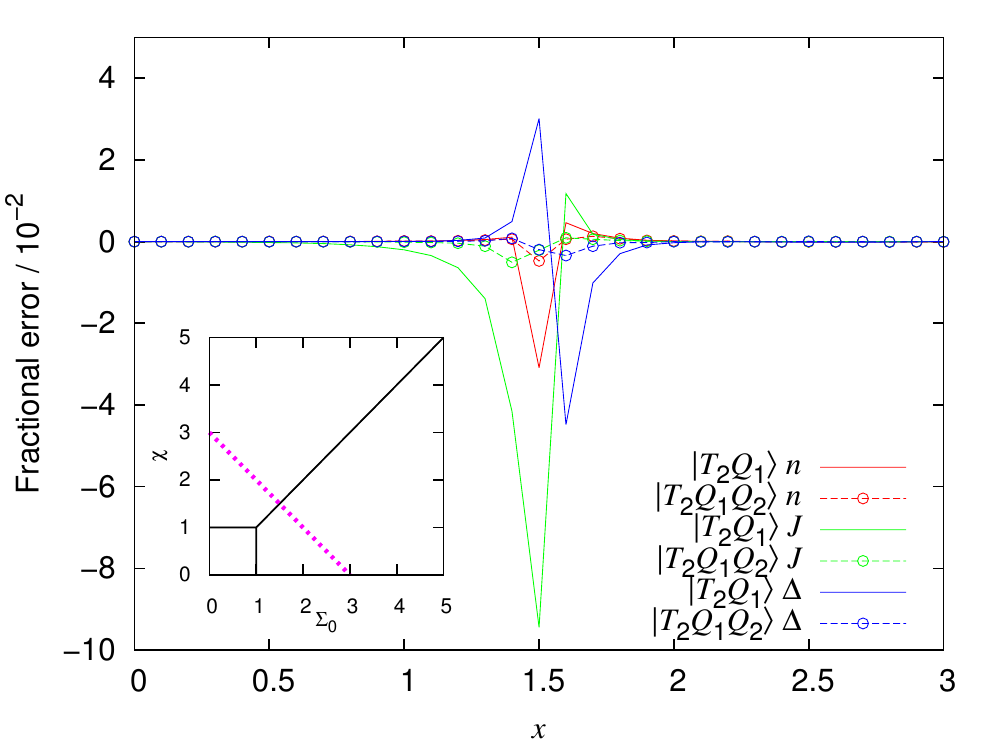}
\caption{$a=3$
\label{OrderParam_error30}}\
\end{subfigure}
\caption{Fractional errors from the exact result of order parameters
for the 40-particle Agassi model obtained from various wave functions, plotted along two lines through
the phase diagram, $\Sigma_0=x$ and $\chi=a-x$ with various $a$. The lines are are displayed in insets.
\label{OrderParam_error}}\
\end{figure}

Holding the above in mind, we examine the errors from FCI of the three above-described order parameters for the vPRCC ($|T_2Q_1\rangle$)
and vPQCC ($|T_2Q_1Q_2\rangle$) methods, which are plotted for two lines through the phase diagram in Fig.\ \ref{OrderParam_error}.
These order parameters are estimated to within 0.5\% in most regions of the phase diagram by both vPRCC and vPQCC. In the weakly correlated
region, the values of all three order parameters are nearly zero;\ in the strongly Lipkin-correlated regions and strongly pairing-correlated
regions, $|T_2Q_1\rangle$ and $|T_2Q_1Q_2\rangle$ display minor errors. However, in the simultaneous strongly Lipkin-correlated and pairing-correlated region
(the center of Fig.\ \ref{OrderParam_error30}), the errors of $|T_2Q_1\rangle$ estimation of all three parameters jump
to 1-10\%. Nevertheless, the $|T_2Q_1Q_2\rangle$ wave function substantially cures this error.
These results further support the inference above that $|T_2Q_1Q_2\rangle$ accounts for coupling between the two
correlation channels which is important when the two are of similar strength and both symmetries are broken - the region where
$|T_2Q_1Q_2\rangle$ is a significant improvement over $|T_2Q_1\rangle$ in the order parameters is the same region where it is an improvement in the energies.

\subsection{The Similarity-Transformation Formalism \label{sec_results_st}}

Although the variational method is robust, it is not practical for realistic systems because
variational coupled cluster equations do not truncate until the order of excitations
reaches the number of particles \cite{Voorhis2000}.
Here we experiment with replacing the variational
formalism with a more standard projective formalism \cite{ShavittBartlett},
in which the Hamiltonian is similarity-transformed with the coupled-cluster
operator and then left-projected against selected excitations.

The Schr\"{o}dinger equation, when approximated with a PRCC or PQCC ansatz, is
\begin{eqnarray}
  He^{T}|Q\rangle &=& Ee^T|Q\rangle,
  \label{Schrodinger_CC}
\end{eqnarray}
where $|Q\rangle$ is the relevant symmetry collective state [Eqs.\ (\ref{nomenclature_def}) and (\ref{nomenclature_example})], i.e.\
PHF ($|Q_1\rangle$) for $|T_2Q_1\rangle$ or $|Q_1Q_2\rangle$ for $|T_2Q_1Q_2\rangle$.
We multiply by $e^{-T}$ on the right to obtain our similarity transformation,
\begin{eqnarray}
  e^{-T}He^{T}|Q\rangle &=& \bar{H}|Q\rangle = E|Q\rangle,
\end{eqnarray}
which renders the transformed Hamiltonian, $\bar{H}$, non-Hermitian.
We project the Schr\"{o}dinger equation against the left-hand eigenstate of $\bar{H}$, which we parameterize as
\begin{eqnarray}
  \langle Q^\prime|(1+Z)\bar{H} &=& E\langle Q^\prime|(1+Z),
\end{eqnarray}
where the prime on $\langle Q^\prime|$ indicates that the $Q$ amplitudes defining the symmetry collective state are, in general,
allowed to differ from those in the right-hand eigenstate, $|Q\rangle$, and
where $Z$ contains de-excitation operators conjugate to the excitations in $T$. For instance, $Z_2$ is
\begin{eqnarray}
  Z_2 &=& z_\mathrm{LL} J_-^2 + z_\mathrm{PP} A_{-1}^\dagger A_{+1}.
\end{eqnarray}
The overall energy expression is
\begin{eqnarray}
  E &=& \frac{\langle Q^\prime|(1+Z)\bar{H}|Q\rangle}{\langle Q^\prime|(1+Z)|Q\rangle},
  \label{ST_energy_eq}
\end{eqnarray}
and we make the energy stationary with respect to all $T$, $Z$, and $Q$ amplitudes.

Note that we have only similarity-transformed with the CC part of our method, and left the symmetry
collective state essentially variational, or rather bivariational, as the bra and ket have different $Q$ amplitudes.
That is, we have not explicitly performed a polynomial similarity transformation \cite{Degroote2016,Qiu2016},
$F^{-1}(Q)HF(Q)$ where $F(Q)$ is defined by $F(Q)|0\rangle=Pe^{Q}|0\rangle$. If we had, it would be necessary to determine
the form of the left-hand eigenstate of the doubly-similarity-transformed Hamiltonian, which is not trivial.
This problem is discussed for the case of spin-projected unrestricted Hartree--Fock (SUHF) in Ref.\ \onlinecite{Qiu2016}.

\begin{figure}
\begin{subfigure}[b]{0.5\textwidth}
\includegraphics[scale=1]{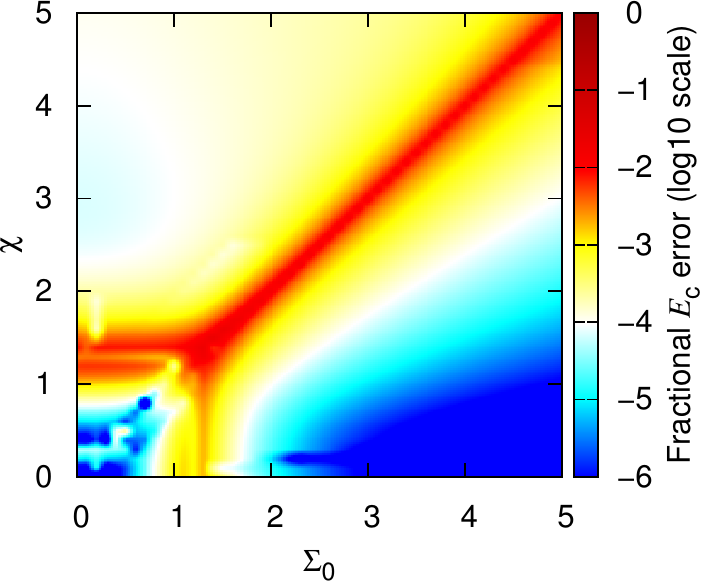}
\caption{PRCC ($|T_2Q_1\rangle$) method.
\label{EPRCCD}}\
\end{subfigure}

\begin{subfigure}[b]{0.5\textwidth}
\includegraphics[scale=1]{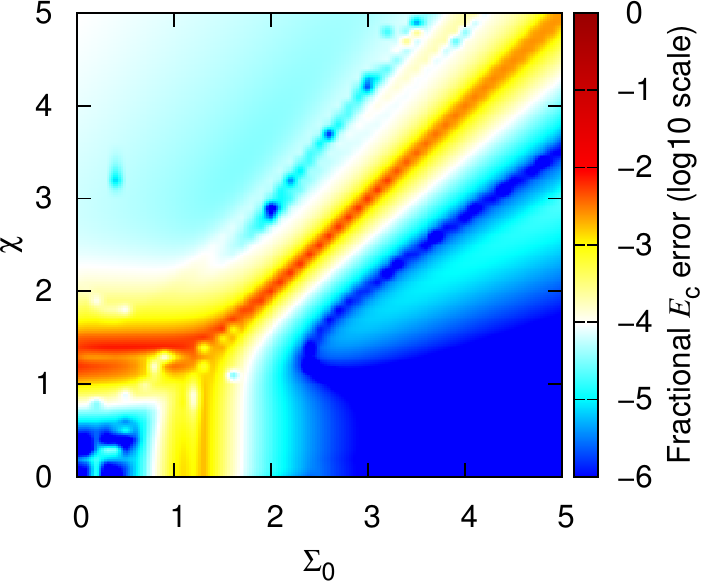}
\caption{PQCC ($|T_2Q_1Q_2\rangle$) method.
\label{EPQCCSd}}\
\end{subfigure}
\caption{
The absolute value of the energy error from FCI for an Agassi model with 40 particles, calculated
with the PRCC ($|T_2Q_1\rangle$) and PQCC ($|T_2Q_1Q_2\rangle$) methods using a similarity-transformation formalism.
\label{EST}}\
\end{figure}

Figure \ref{EST} shows the error from FCI of the PRCC ($|T_2Q_1\rangle$) and PQCC ($|T_2Q_1Q_2\rangle$) methods, respectively. The accuracy is slightly
reduced in both cases from the variational results, but not qualitatively different. $|T_2Q_1\rangle$ is extremely accurate in all except
recoupling regions and diagonal strong correlation, and $|T_2Q_1Q_2\rangle$ remedies its defect in the latter, with correlation energies accurate
to within 1\%. These results confirm that the combination of coupled cluster with symmetry collective states
improves dramatically on the results demonstrated in Fig.\ \ref{RCCD_breakdown} in Sec.\ \ref{sec_theory_agassi};\ at every point on the phase
diagram, real solutions to the equations for the $T$ amplitudes are obtained, despite the fact that we have discarded the variational
method.

\section{Conclusion \label{sec_conc}}

The PRCC and PQCC methods, initially tested on the Lipkin model Hamiltonian \cite{Wahlen2017}, are shown to very effectively
estimate the wave function of the Agassi model, recovering more than 99\% of correlation energy and predicting
order parameters of the wave function with very high accuracy. While the previous work \cite{Wahlen2017} demonstrated the principle by which
such methods could model systems under strong correlation, here we have shown how they can be used for systems with multiple
distinct correlation channels, including multiple simultaneous motifs of strong correlation.
Certainly, more work is necessary in generalizing this formalism to realistic nuclear and electronic structure problems
of nuclear physics, quantum chemistry, and condensed matter physics. However,
the fact that methods with between 5 and 8 parameters including no excitation amplitudes beyond the level of doubles are so powerful
for this simple model system is very promising about the potential of symmetry-collective-state-plus-CC approaches to strong correlation.

\begin{acknowledgments}
We thank Jacob M.\ Wahlen-Strothman and Dr.\ Matthias Degroote for providing reference data for testing purposes, and Dr.\ Thomas M.\ Henderson for assistance
in overcoming computational problems.
This work was supported by the U.S. Department of Energy, Office of Basic Energy Sciences, Computational and Theoretical Chemistry Program under Award No. DE-FG02-09ER16053. G.E.S. is a Welch Foundation Chair (No. C-0036). Computational resources for this work were supported in part by the Big-Data Private-Cloud Research Cyberinfrastructure MRI-award funded by NSF under grant CNS-1338099 and by Rice University. J.D. acknowledges support from the Spanish Ministry of Economy and Competitiveness and FEDER through Grant No. FIS2015-63770-P.
\end{acknowledgments}

\appendix

\section{Algebra of the Agassi model \label{sec_appendix_O5}}

\begin{table*}[tb]
\begin{ruledtabular}
\caption{Commutators of the 10 generators of the O(5) algebra of the Agassi model, in the format [row, column]. The mappings of all operators to fermions
are given in Eqs.\ (\ref{Agassi_Operators_1}), (\ref{number_op}), (\ref{Agassi_Operators_2}), and (\ref{Agassi_Operators_derived}), and $j$ is 1/2 times the number of particles
in the system.
\label{Agassi_commutators}}
\begin{tabular}{l||c|c|c|c|c|c|c|c|c|c}
			& $J_+$			& $J_-$			& $J_0$			& $A_{+1}^\dagger$	& $A_{+1}$	& $A_{-1}^\dagger$	& $A_{-1}$	& $A_0^\dagger$		& $A_0$		& $N$	\\ \hline
 $J_+$			& 0			& $2J_0$		& $-J_+$		& 0			& $-A_0$	& $A_0^\dagger$		& 0		 & $2A_{+1}^\dagger$	& $-2A_{-1}$	& 0	\\
 $J_-$			& $-2J_0$		& 0			& $J_-$			& $A_0^\dagger$		& 0		& 0			& $-A_0$	 & $2A_{-1}^\dagger$	& $-2A_{+1}$	& 0	\\
 $J_0$			& $J_+$			& $-J_-$		& 0			& $A_{+1}^\dagger$	& $-A_{+1}$	& $-A_{-1}^\dagger$	& $A_{-1}$	& 0			& 0 		& 0	\\
 $A_{+1}^\dagger$	& 0			& $-A_0^\dagger$	& $-A_{+1}^\dagger$	& 0			& $(N_{+1}-j)$	& 0			 & 0		& 0			& $J_+$		& $-2A_{+1}^\dagger$	\\
 $A_{+1}$		& $A_0$			& 0			& $A_{+1}$		& $(j-N_{+1})$		& 0		& 0			& 0		& $-J_-$		& 0		& $2A_{+1}$	\\
 $A_{-1}^\dagger$	& $-A_0^\dagger$	& 0			& $A_{-1}^\dagger$	& 0			& 0		& 0			& $(N_{-1}-j)$	& 0			& $J_-$		& $-2A_{-1}^\dagger$	\\
 $A_{-1}$		& 0			& $A_0$			& $-A_{1}$		& 0			& 0		& $(j-N_{-1})$		& 0		& $-J_+$		& 0		& $2A_{-1}$	\\
 $A_0^\dagger$		& $-2A_{+1}^\dagger$	& $-2A_{-1}^\dagger$	& 0			& 0			& $J_-$		& 0			 & $J_+$		& 0			& $(N-2j)$	& $-2A_0^\dagger$	\\
 $A_0$			& $2A_{-1}$		& $2A_{+1}$		& 0			& $-J_+$		& 0		& $-J_-$		& 0		& $(2j-N)$		& 0 		& $2A_0$ \\
 $N$			& 0			& 0			& 0			& $2A_{+1}^\dagger$	& $-2A_{+1}$	& $2A_{-1}^\dagger$	& $-2A_{-1}$	& $2A_0^\dagger$	& $-2A_0$	& 0
\end{tabular}%
\end{ruledtabular}
\end{table*}%

The Hamiltonian of the Agassi model [Eq.\ (\ref{Agassi_Hamiltonian})] is given in Sec.\ \ref{sec_theory_agassi} terms of seven
one-body operators, whose mapping to fermions is given in Eqs.\ (\ref{Agassi_Operators_1}). If we add to these seven
the number operator [$N$, Eq.\ (\ref{number_op})] along with
$A_0^\dagger$ ($A_0$), which creates (destroys) a particle $\sigma=+1$ and another one in $\sigma=-1$,
\begin{subequations}
\begin{eqnarray}
  A_0^\dagger &=& \sum_{m>0} \left( c_{-1,m}^\dagger c_{+1,-m}^\dagger - c_{-1,-m}^\dagger c_{+1,m}^\dagger\right),
  \\
  A_0 &=& \sum_{m>0} \left( c_{+1,-m} c_{-1,m} - c_{+1,m} c_{-1,-m} \right),
\end{eqnarray}
\label{Agassi_Operators_2}
\end{subequations}
then the combined set of ten one-body operators closes an O(5) algebra.
All commutators among the ten $A$, $J$, and $N$ operators
are summarized in Table \ref{Agassi_commutators}.
Note that the SU(2) algebra of the Lipkin model 
and the three SU(2) algebras of the pairing model are subalgebras of this O(5) \cite{Klein1982}.
Other useful one-body operators can be defined as linear combinations of these
ten generators; for instance, the number of particles in the upper and lower levels are given by
\begin{subequations}
\begin{eqnarray}
  N_{+1} &=& N/2 + J_0
  \nonumber\\ &=& \sum_m c_{+1,m}^\dagger c_{+1,m},
  \\
  N_{-1} &=& N/2 - J_0
  \nonumber\\ &=& \sum_m c_{-1,m}^\dagger c_{-1,m}.
\end{eqnarray}
\label{Agassi_Operators_derived}
\end{subequations}

Following Ref.\ \cite{Klein1982}, a complete (although non-orthogonal) set of states for the Agassi model can likewise be defined in terms
of pair creation ($A^\dagger$) operators acting on the physical vacuum:
\begin{eqnarray}
  |n_-,n_+,n_0\rangle &=& \left( A_{-1}^\dagger \right)^{n_-}
                  \left( A_{+1}^\dagger \right)^{n_+}
                  \left( A_0^\dagger    \right)^{n_0} |-\rangle,
  \label{Agassi_States}
\end{eqnarray}
which is complete because all other generators in the $O(5)$ algebra annihilate the physical
vacuum to the right. Furthermore, if we seek only the states with $2j$ particles, then
we have $n_-+n_++n_0 = j$ and the size of the Hilbert space is therefore quadratic in $j$, as asserted
in Sec.\ \ref{sec_theory_agassi}. As for parity symmetry, even or odd sectors correspond simply to even or odd $(n_0-j)$.
The symmetry-adapted non-interacting ground state ($|0\rangle$ in the main text) in these terms is
$|j,0,0\rangle$.

By repeatedly applying the commutation relationships summarized in Table \ref{Agassi_commutators}, the effects
of all ten generators on a state $|n_-,n_+,n_0\rangle$ can be evaluated:
\begin{widetext}
\begin{subequations}
\begin{eqnarray}
  J_+|n_-,n_+,n_0\rangle &=& n_-|n_--1,n_+,n_0+1\rangle
  + 2n_0|n_-,n_++1,n_0-1\rangle,
  \label{Jp_Ket}
  \\
  J_-|n_-,n_+,n_0\rangle &=& n_+|n_-,n_+-1,n_0+1\rangle
  + 2n_0|n_-+1,n_+,n_0-1\rangle,
  \label{Jm_Ket}
  \\
  J_0|n_-,n_+,n_0\rangle &=& (n_+-n_-)|n_-,n_+,n_0\rangle,
  \label{J0_Ket}
  \\
  A_{+1}^\dagger|n_-,n_+,n_0\rangle &=& |n_-,n_++1,n_0\rangle,
  \label{Apd_Ket}
  \\
  A_{+1}|n_-,n_+,n_0\rangle &=& n_+(j-n_++1-n_0)|n_-,n_+-1,n_0\rangle
  - n_0(n_0-1)|n_-+1,n_+,n_0-2\rangle,
  \label{Ap_Ket}
  \\
  A_{-1}^\dagger|n_-,n_+,n_0\rangle &=& |n_-+1,n_+,n_0\rangle,
  \label{Amd_Ket}
  \\
  A_{-1}|n_-,n_+,n_0\rangle &=& n_-(j-n_-+1-n_0)|n_--1,n_+,n_0\rangle
  - n_0(n_0-1)|n_-,n_++1,n_0-2\rangle,
  \label{Am_Ket}
  \\
  A_0^\dagger|n_-,n_+,n_0\rangle &=& |n_-,n_+,n_0+1\rangle,
  \label{A0d_Ket}
  \\
  A_0|n_-,n_+,n_0\rangle &=& n_0(2j-n_0-2n_--2n_++1)|n_-,n_+,n_0-1\rangle
  - n_-n_+|n_--1,n_+-1,n_0+1\rangle,
  \label{A0_Ket}
  \\
  N|n_-,n_+,n_0\rangle &=& 2(n_-+n_++n_0)|n_-,n_+,n_0\rangle.
  \label{N_Ket}
\end{eqnarray}
\label{Op_Ket_relations}
\end{subequations}
\end{widetext}
Equations (\ref{Ap_Ket}), (\ref{Am_Ket}), and (\ref{A0_Ket}) can be utilized to construct recursive equations
that are used to evaluate the overlap matrix of the states $|n_-,n_+,n_0\rangle$.
Keeping in mind that states with differing $N$ and $J_0$ eigenvalues are orthogonal,
and that $\langle 0,0,0|0,0,0\rangle = \langle-|-\rangle = 1$, we have
\begin{widetext}
\begin{subequations}
\begin{eqnarray}
  \langle n_-^\prime,   n_+^\prime,   n_0^\prime   |n_-,   n_+,   n_0   \rangle
  &=& n_- (j-n_-+1-n_0)
  \langle n_-^\prime-1, n_+^\prime,   n_0^\prime   |n_--1, n_+,   n_0   \rangle
  \nonumber \\ && -n_0(n_0-1)
  \langle n_-^\prime-1, n_+^\prime,   n_0^\prime   |n_-,   n_++1, n_0-2 \rangle,
  \\ &=& n_+ (j-n_++1-n_0)
  \langle n_-^\prime,   n_+^\prime-1, n_0^\prime   |n_-,   n_+-1, n_0   \rangle
  \nonumber \\ && -n_0(n_0-1)
  \langle n_-^\prime,   n_+^\prime-1, n_0^\prime   |n_-+1, n_+,   n_0-2 \rangle,
  \\ &=& n_0 (2j-n_0-2n_--2n_++1)
  \langle n_-^\prime,   n_+^\prime,   n_0^\prime-1 |n_-,   n_+,   n_0-1 \rangle
  \nonumber \\ && -n_-n_+
  \langle n_-^\prime,   n_+^\prime,   n_0^\prime-1 |n_--1, n_+-1, n_0+1 \rangle.
\end{eqnarray}
\label{Agassi_Overlap}
\end{subequations}
\end{widetext}

Given the overlap matrix [Eqs.\ (\ref{Agassi_Overlap})] and the actions of all ten generators on
the non-orthogonal basis [Eqs.\ (\ref{Op_Ket_relations})], an orthonormal
basis and matrix elements of all ten operators are easily obtained. Since, as demonstrated,
the size of the basis (at least in the symmetry-adapted space) grows quadratically
with $j$, the overlap and (symmetry-adapted products of) operator matrices in the symmetry-adapted space
require only $O(j^4)$ storage space. The largest computational difficulty is numerical instability
in systems of $10^2$ particles or more, arising from overlap matrix elements with values that differ by more than 16 orders
of magnitude, which is partially resolved by using quadruple-precision storage of floating-point numbers.
This means that FCI and all of the symmetry-adapted methods
described in this work are easily implemented using complete-basis representations of all operators.
The symmetry-broken mean-field is another matter,
however this can be calculated very easily using simple closed-form equations as explained by Davis and Heiss \cite{Davis1986}.

\bibliographystyle{apsrev}
\bibliography{library}
\end{document}